\documentclass[fleqn,usenatbib]{mnras}

\usepackage{newtxtext,newtxmath}

\usepackage[T1]{fontenc}
\usepackage{ae,aecompl}
\DeclareRobustCommand{\VAN}[3]{#2}
\let\VANthebibliography\thebibliography
\def\thebibliography{\DeclareRobustCommand{\VAN}[3]{##3}\VANthebibliography}

\usepackage{graphicx}	
\usepackage{amsmath}	
\usepackage{geometry}
\usepackage{ulem}
\usepackage{tabularx,booktabs}
\usepackage{float}
\usepackage[justification=centering]{caption}
\newcommand{\pcm}{\,cm$^{-2}$}	
\newcommand{\ps}{\,s$^{-1}$}	
\newcommand{\cps}{\,counts\,s$^{-1}$}	
\newcommand{\pcms}{\,cm$^{-2}$\,s$^{-1}$}	
\newcommand{\angstrom}{\textup{\AA}} 

\title[\textit{AstroSat}, \textit{NuSTAR} and \textit{Swift} observations of 4U~1957+115]{Spectral characteristics of the black hole binary 4U~1957+115:\\ A multi-mission perspective}

\author[Mudambi et al.]{
S. P. Mudambi,$^{1}$
S. B. Gudennavar,$^{1}$ \thanks{E-mail: shivappa.b.gudennavar@christuniversity.in; sneha.m@res.christuniversity.in; rmsira@iucaa.in}
R. Misra $^{2}$
and S. G. Bubbly$^{1}$
\\\\
$^{1}$Department of Physics and Electronics, CHRIST (Deemed to be University), Bangalore Central Campus, Bengaluru-560029, India\\
$^{2}$Inter-University Centre for Astronomy and Astrophysics, Ganeshkind, Pune-411007, India
}

\date{Accepted XXX. Received YYY; in original form ZZZ}

\pubyear{2021}

\begin{document}
\label{firstpage}
\pagerange{\pageref{firstpage}--\pageref{lastpage}}
\maketitle
\begin{abstract}
We report spectral analysis of the persistent black hole X-ray binary, 4U 1957+115, using \textit{AstroSat, Swift} and \textit{NuSTAR} observations carried out between 2016-2019. Modelling with a disk emission, thermal comptonization and blurred reflection components revealed that the source was in the high soft state with the disk flux $\sim 87$\% of the total and high energy photon index  $\sim2.6$. There is an evidence that either the inner disk radius varied by $\sim25$\% or the colour hardening factor changed by $\sim12$\%. The values of the inner disk radius imply that for a non-spinning black hole, the black hole mass is $<7$~M$_\odot$ and the source is located $>30 $ kpc away. On the other hand, a rapidly spinning black hole would be consistent with the more plausible black hole mass of $< 10$~M$_\odot$ and a source distance of $\sim10$ kpc. Fixing the distance to $10$ kpc and using a relativistic accretion disk model, constrained the black hole mass to 6~M$_\odot$ and inclination angle to 72$^{\circ}$. A positive correlation is detected between the accretion rate and inner radii or equivalently between the accretion rate and colour factor.
\end{abstract}
\section{Introduction}
4U~1957+115 is a black hole low mass X-ray binary (BH-LMXB) that was diskovered by \textit{Uhuru} in 1973 \citep{Gia74} and its optical counterpart V1408 Aql was identified four years later during the Aquila survey \citep{Mar78}. 
\citet{Tho87} estimated the orbital period of it to be $9.329\pm0.011$~hr by studying the photometric modulations in the optical lightcurves, probably due to the X-ray irradiation of the donor by the accretion disk and the accretor. 
\\[6pt]
\citet{Hak99} using the simultaneous data from UBVRI filters of the \textit{Nordic Optical Telescope} at La Palma reported a high disk inclination angle ($i = 70-75^{\circ}$). Furthermore, they suggested that optical modulations are due to the changes in accretion disk structure but not due to the changes in X-ray heated face of the donor. \citet{Bay11} and \citet{Mas12} analysed the high-speed optical photometry data from Argos CCD photometer mounted on the $2.$1~m \textit{Otto Struve Telescope} at McDonald Observatory and provided a weak constraint on the orbital inclinations ($20^{\circ}<~i~<70^{\circ}$) and mass ratios ($0.025-0.3$). \citet{Gom15} constrained the black hole mass ($M_{BH}$) and orbital inclination angle to be $3$~M$_{\odot}<$~$M_{BH}$~<~6.2~M$_{\odot}$ and $12.75^{\circ}$, respectively using Argos CCD photometer data. They found the source distance to lie between $20-40$~kpc.
\\[6pt]
\citet{Yaq93} analysed seventeen \textit{GINGA} observations ($1.0-18.0$~keV) of the source and estimated the $\mathtt{powerlaw}$ index to be $2-3$. They found the spectra to be dominated by a stable soft disk component and a varying hard $\mathtt{powerlaw}$ component. Spectral analysis of \textit{ASCA} ($0.5-10.0$~keV) and \textit{RXTE/PCA} ($0.8-10.0$~keV) data hinted the presence of a highly inclined accretion disk ($i$~$>65^{\circ}$). \textit{RXTE/ASM} ($2.0-20.0$~keV) data revealed a $117$~day period likely due to a warped precessing accretion disk \citep{Now99}. However, a more detailed study of the entire \textit{RXTE/ASM} data showed that the X-ray periodicity of 4U~1957+115 is variable and not exactly $117$~days as predicted earlier and is probably due to the variations in the mass accretion rate on to the black hole \citep{Wij02}. \citet{Now08} using data from \textit{RXTE/PCA} ($3.0-18.0$~keV) tried to estimate the black hole spin ($a$) by modelling the disk continuum with $\mathtt{kerrbb}$ \citep{Li05}, $\mathtt{comptt}$ \citep{Tit94} and $\mathtt{gaussian}$ models. X-ray data analysis suggested a rapidly spinning accretor with spin varying in the range $a = 0.83 - 1$. \citet{Now08} suggested that 4U~1957+115 is at a distance greater than $5$~kpc using \textit{Chandra/HETGS} ($0.7-8.0$~keV) and \textit{XMM-Newton} ($0.3-12.0$~keV) data. They probed the equivalent width of the Ne IX $13.45~\angstrom$ line associated with warm/hot phase in the interstellar medium to arrive at this conclusion which was later confirmed by \citet{Yao08}.
\\[6pt]
\citet{Now12} used $\mathtt{diskbb}$ \citep{Mit84,Mak86}, $\mathtt{powerlaw}$, $\mathtt{eqpair}$ \citep{Cop00} and $\mathtt{diskpn}$ \citep{Gie99} models to fit the \textit{Suzaku/XIS} spectrum in the energy range $0.5-8.0$~keV. They found the source spectra to have limited hard photon excess (E~$>8.0$~keV). For a set of mass and distance values (i.e. $3$~M$_{\odot}$ and $10$~kpc and  $16$~M$_{\odot}$ and $22$~kpc), the black hole spin ($a$) was estimated to be greater than $0.9$ for an assumed disk inclination angle of $75^{\circ}$. \citet{Mai14} analysed 26 data sets from \textit{Swift/XRT} ($0.3-8.0$~keV) and estimated the disk inclination angle to be $\sim$77.6$^{\circ}$ using Monte Carlo simulations and chi-square fitting method. They modelled the spectra with $\mathtt{kerrbb}$ using several combinations of mass, inclination angle and distance triplets. Their results indicated a maximally spinning black hole ($a\sim0.98$). The colour hardening factor was found to be slightly higher ($1.9-2.1$) than the typical value of $1.7$ \citep{Shi95}. \citet{Sha21} analysed nine \textit{NuSTAR} data sets in the energy range $3.0-40.0$~keV and attempted to constrain the black hole mass and spin. The spectra was modelled primarily using $\mathtt{kerrbb}$ along with $\mathtt{thcomp}$ \citep{Zdz20} and $\mathtt{gaussian}$ models. 
Their results inferred a moderate spin of $0.85$ for a black hole mass of $4-6$~M$_{\odot}$ at a source distance of $7$~kpc. 
\\[6pt]
While a canonical black hole X-ray binary alternates between hard and soft spectral states, 4U~1957+115 has always been observed in the high soft state with flux levels between $20$ and $70$~mCrab ($2.0-12.0$~keV) \citep{Yaq93,Now99,Wij02,Now08,Dun10,Mai14}. Despite being one of the most studied sources by various X-ray missions, precise values of the black hole spin, mass and distance of 4U 1957+115 are yet to be determined. Most of the accepted estimates are obtained primarily from RXTE data which was sensitive to energy $>3$~keV while the characteristic temperature of the source is $\sim$1~keV \citep{Now99}. Indeed, the spectral parameters of the BH-LMXB 4U 1957+115 in the soft state can be better constrained if it is observed by an instrument with broadband spectral coverage with the sensitivity $<1$~keV. Data from Soft X-ray Telescope (SXT) \citep{Sin16,Sin17} and Large Area X-ray Proportional Counter (LAXPC) \citep{Yad16,Agr17} aboard the Indian Space Observatory \textit{AstroSat} \citep{Agr06,Sin14,Agr17} provide such an opportunity with simultaneous broadband spectral coverage in the energy range $0.3-80.0$~keV. Furthermore, there is also simultaneous data available from \textit{Swift/}XRT ($0.3-7.0$~keV) \citep{Bur05} and  \textit{NuSTAR/}FPMA and \textit{NuSTAR/}FPMB ($4.0-79.0$~keV) \citep{Har2013} to study the evolution of spectral parameters of 4U 1957+115.
\\[6pt]
In this work, we attempt to constrain the spectral parameters of 4U 1957+115 via comprehensive broadband spectral analysis using simultaneous data from SXT and LAXPC of \textit{AstroSat}, XRT of \textit{Swift} and FPMA and FPMB of \textit{NuSTAR}. The paper is structured as follows. First, we discuss the data reduction and extraction procedure in \S~$2$. Next, we present the steps of broadband spectral analysis in \S~$3$. Lastly, we discuss our results and summarize them in \S~$4$.
\section{Observations and Data reduction}\label{datred}
\subsection{\textit{AstroSat}/\textbf{SXT} and \textbf{LAXPC}}
\begin{figure}
\centering
\includegraphics[width=0.34\textwidth,angle=-90]{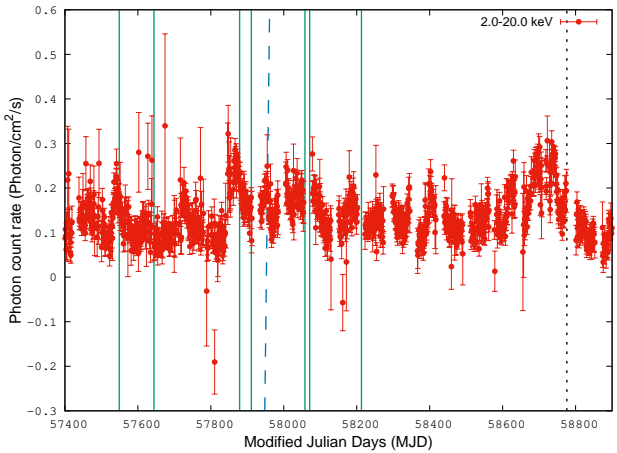}
\caption{$2.0-20.0$~keV \textit{MAXI} lightcurve of 4U~1957+115. The vertical lines green (solid), blue (dashed) and black (dotted) shows \textit{AstroSat}, \textit{Swift} and \textit{NuSTAR} observations of the source considered for the study respectively.}
\label{maxilc}
\end{figure}

The X-ray instruments SXT and LAXPC  observed the source on several occasions between $2016$ and $2018$ (Figure \ref{maxilc}), which are referred to as Epochs (Table \ref{obstab}). Level~$2$ data of SXT was deduced from Level~$1$ data using AS1SXTLevel2-1.4b\footnote{\url{http://www.tifr.res.in/~astrosat_sxt/sxtpipeline.html}} version of \textsc{sxt} pipeline. \textsc{xselect} $V2.4k$ tool of \textsc{heasoft} ($v6.28$) was used to extract individual SXT spectra from their respective processed Level~$2$ event files. Each of these spectra was extracted from a circular region of $16^\prime$ radius centred on the source. The net countrate in the energy range $0.3-7.0$~keV was found to be $16$~\cps. Therefore, an exclusion of the central source region of $1-3^\prime$ radius to account for the pile-up was deemed not necessary. 
Auxiliary response file (ARF) ``corr$\_$crab$\_$long2018$\_$616A.arf'', response matrix file (RMF) ``~sxt$\_$pc$\_$mat$\_$g0to12.rmf~'' and background spectrum ``~SkyBkg$\_$comb$\_$EL3p5$\_$Cl$\_$Rd16p0$\_$v01.pha~'' provided by Payload Operations Centre (POC) were used for the analysis. 
\\[6pt]
Level~$1$ event mode data of LAXPC was converted to Level~$2$ data using the official \textsc{laxpc} pipeline\footnote{\url{http://astrosat-ssc.iucaa.in/?q=laxpcData}}. \textsc{laxpc} command  `laxpc$\_$make$\_$stdgti' was used to generate good time interval file to remove Earth occultation of the source and South Atlantic Anomaly passes of the satellite. Total spectrum along with RMF for LAXPC-10, LAXPC-20 and LAXPC-30 were obtained using `laxpc$\_$make$\_$spectra' subroutine. Background spectrum was extracted by adopting the faint background model \citep{Mis21} of \textsc{laxpc} software as the source has very low countrate in the energy range greater than $30.0$~keV. Since LAXPC-20 had lower background as compared to other two LAXPCs, we used LAXPC-20 data only for this study.
\\[6pt]
The LAXPC data above $20.0$~keV was not considered for the analysis as LAXPC spectrum in the higher energies was found to be background dominated. \textsc{ftools} subroutine \textit{ftgrouppha} with group type set to ``opt'' \citep{Kaa16} was used to bin the SXT spectrum for better statistics. SXT spectrum was considered only from $0.7$ to $6.0$~keV as the instrument has uncertainties in the effective area and response below $<0.7$~keV. A systematic error of 3 \% was added as prescribed by the POC \citep{Ant17,Bhat17}. An offset gain correction of $25$~eV determined with slope fixed at unity was applied to SXT data. Lower and higher energy range were extended to 0.1~keV and 100.0~keV respectively to supply an energy-binning array for the spectral fits.
\begin{table*}
\centering
\caption{\textit{AstroSat}, \textit{Swift} and \textit{NuSTAR} observations of 4U~1957+115}
\begin{center}
\begin{tabular}{c l c c c c c l}
\toprule
Epoch & Obs ID & MJD & Start Time (hh:mm:ss) & End Time (hh:mm:ss)	& Exposure Time (ks)\\
 & & & Date (dd-mm-yyyy) & Date (dd-mm-yyyy) & SXT$~~~$LAXPC \\
\midrule
1 & 9000000490 & 57549 & 22:31:52 & 01:00:19 &  12.93$~~~~~~~$5.10 \\
	   & & & 10-06-2016 & 11-06-2016 &  \\
	   \\
	    & & & 13-09-2016 & 13-09-2016 &	\\
	    \\
2 & 9000001270 & 57879 & 14:09:38 & 18:17:38 & 6.44$~~~~~~~$9.33\\
	 &   & & 05-06-2017 & 05-06-2017 & \\
	 \\
3 & 9000001362 & 57911 & 13:18:26 & 18:15:30 & 6.94$~~~~~~~$9.20 \\
	&   & & 06-07-2017 & 06-07-2017 & \\
	\\
\\
4 & 9000001404  & 57961 & 08:08:13 & 09:13:42 & 7.74$~~~~~~~$18.40\\
& & & 27-07-2017 & 27-07-2017 & \\
\\
5 & 9000001658  & 58058 & 06:34:25 & 11:29:14 & 3.84$~~~~~~~$10.00\\
	   & & & 01-11-2017 & 01-11-2017 & \\
	   \\
6 & 9000002014  & 58213 & 06:18:08 & 06:32:38 & 1.08$~~~~~~~$7.85\\
	    & & & 05-04-2018 & 05-04-2018 & \\
	    \\
7 & 9000001380  & 57948 & 06:52:42 & 08:57:47 & 7.73$~~~~~~~$14.30\\
& & & 14-07-2017 & 14-07-2017 & \\
& 00030959007~$^{*}$  & 57948 & 01:10:40 & 17:10:56 & $--~~~~~~~--$\\ 
& & & 14-07-2017 & 15-07-2017 & \\
\\
8 &  000889750038~$^{*}$  &  58776 & 14:28:26 & 14:56:56 & $--~~~~~~~--$\\
	    & & & 10-20-2019 & 10-20-2019 & \\
  &  30502007010~$^{**}$  &  58776 & 12:51:09 & 23:46:09 & $--~~~~~~~--$\\
	    & & & 10-20-2019 & 10-20-2019 & \\
\bottomrule
$^{*}$\textit{Swift};~~$^{**}$\textit{NuSTAR} and \textit{Swift}
\end{tabular}
\end{center}
\label{obstab}
\end{table*}
\subsection{\textit{Swift/}XRT}
It was found that both \textit{AstroSat} and \textit{Swift} have observed 4U~1957+115 simultaneously on 2017, July 07 (MJD~$57948$). Motivated by this, we proceeded with the joint spectral analysis of the data from \textit{Swift/}XRT and SXT, LAXPC-20 of \textit{AstroSat}. \textit{Swift} data extraction and reduction were performed as per the steps outlined in \citet{Rey13} using \textsc{heasoft} software ($v6.28$). \textsc{xrtpipeline} command was used to reprocess the raw windowed timing mode data and then to apply the latest instrument calibrations and responses. The \textit{Swift} spectra of the source were extracted from a circular region of $47^{\prime \prime}$ centred around the source image. Adjoining source free region of the same size was used to extract the background spectrum. We used the latest RMF provided by the POC team and an ARF file created using \textsc{xrtmkarf} task for the analysis. \textit{Swift} spectra were also binned on the similar lines of SXT spectra for better statistics. Spectral analysis was carried out in the energy range $0.3-6.0$~keV. A 3~\% systematic error, as mentioned in \textit{Swift}/XRT CALDB Release Note4, was added while fitting. 
\subsection{\textit{NuSTAR/}FPMA and FPMB}
Of the many \textit{NuSTAR} observations of the source, 4U 1957+115 was observed simultaneously by both \textit{Swift} and \textit{NuSTAR} (FPMA and FPMB telescopes) on 2019, October 20 (MJD~$58776$). The \textit{NuSTAR} data were then processed using the latest \textsc{nustardas} software in the \textsc{heasoft} ($v6.28$) along with the latest calibration files (CALDB). The \textsc{nupipeline} (version $0.4.8$) task was used for filtering event files. Total spectrum was extracted from a circular region of $10^\prime$ radius centred around the source whereas the background spectrum was extracted from a circular region of the same size farthest from the source. The nuproducts task was used to generate RMF and ARF for both telescopes- FPMA and FPMB. Total spectrum in the energy range $4.0-50.0$~keV was considered for the study. No systematic error was added during the analysis.
\section{Spectral analysis}
\label{spectra}
\begin{figure}
\centering
\includegraphics[width=0.34\textwidth,angle=-90]{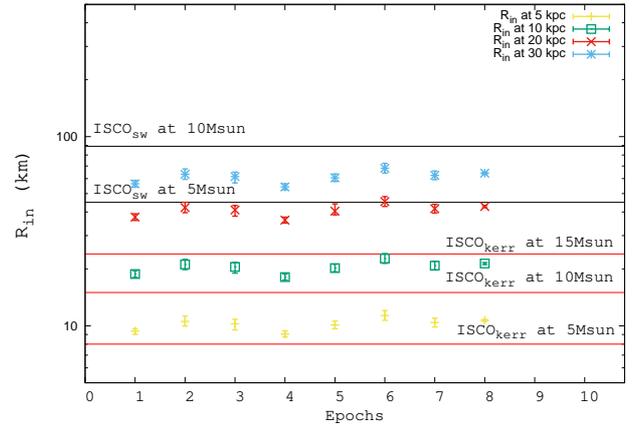}
\caption{Absolute inner disk radius (R$_{in}$) as a function of Epochs calculated at $5$~kpc, $10$~kpc, $20$~kpc and $30$~kpc. Each of the horizontal lines represents $R_{\rm{ISCO}}$ values for maximally spinning (ISCO$_{kerr}$) black hole of mass $5$~M$_{\odot}$, $10$~M$_{\odot}$ and $15$~M$_{\odot}$ and non-spinning (ISCO$_{sw}$) black hole of mass $5$~M$_{\odot}$ and $10$~M$_{\odot}$.}
\label{rin}
\end{figure}

\begin{figure*}
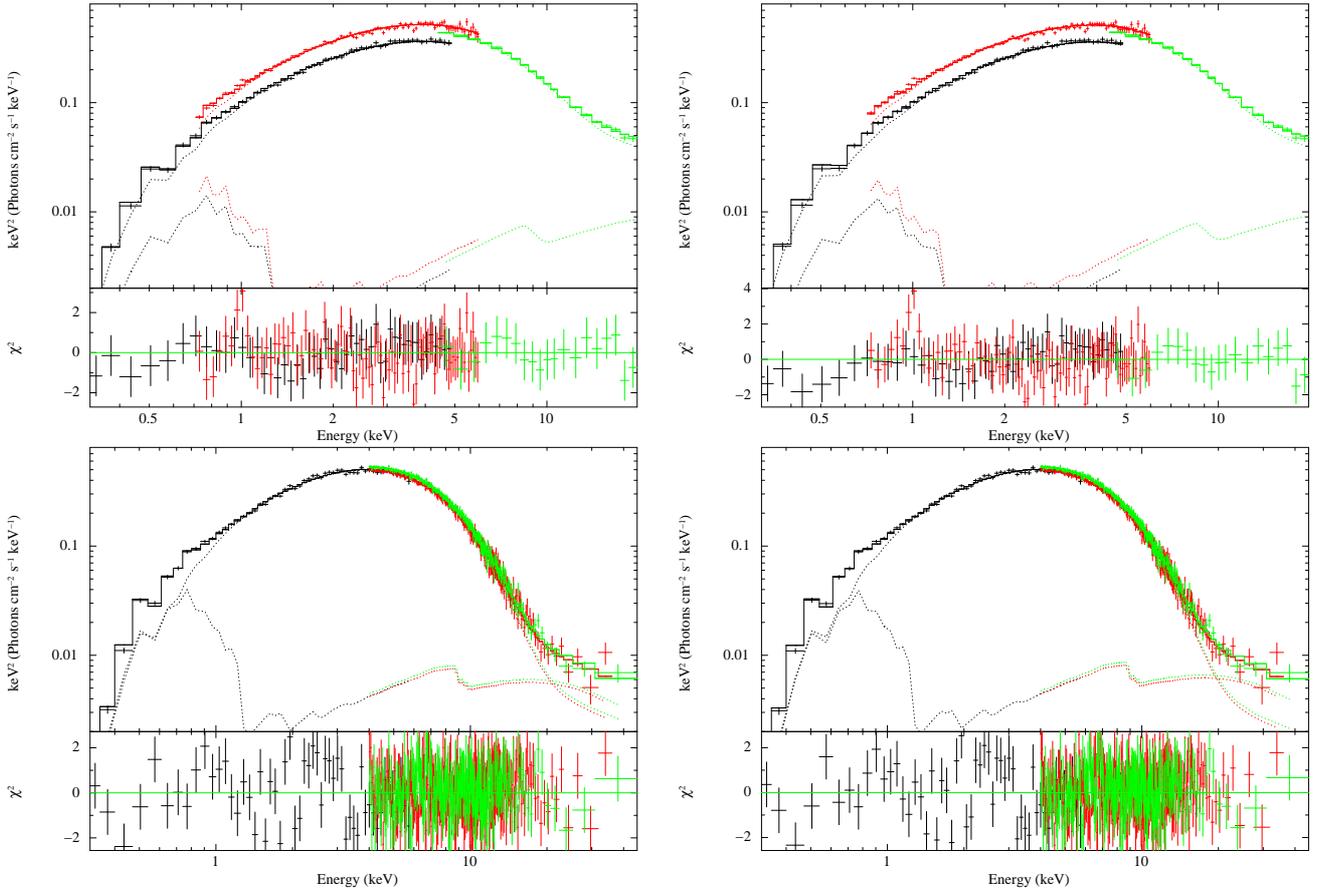

    \centering
    \includegraphics[width=0.33\textwidth,angle=-90]{figures/swiastdisk_grp.eps}
    \includegraphics[width=0.33\textwidth,angle=-90]{figures/swiastkerrd_grp.eps}
     \includegraphics[width=0.33\textwidth,angle=-90]{figures/swinudisk_grp.eps}
    \includegraphics[width=0.33\textwidth,angle=-90]{figures/swinukerrd_grp.eps}
    \caption{The combined spectra for Epoch 7 (Top panels) and Epoch 8 (Bottom panels). The left panels show the best fitted model using the non-relativistic disk model ($\mathtt{constant\times~tbabs\times(relxill+simpl\otimes~diskbb)}$), while the right panels show the fitting using the relativistic disk model ($\mathtt{constant\times~tbabs\times(relxill+simpl\otimes~kerrd)}$).}
    \label{combspec}
\end{figure*}

\begin{figure}
\includegraphics[width=0.50\textwidth]{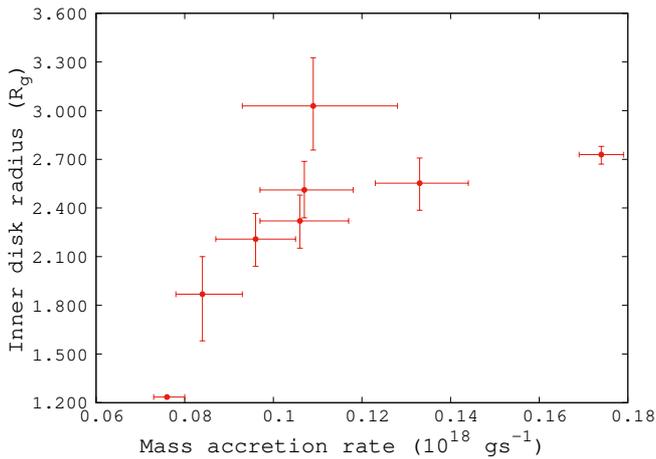}
\caption{Inner disk radius in the units of R$_g$ of the source as function of mass accretion rate for all the eight Epochs.}
\label{parvar}
\end{figure}

We carried out the combined spectral analysis of the data from \textit{AstroSat}/SXT$-$LAXPC-20, \textit{Swift} and \textit{NuSTAR} observations using \textsc{xspec} version$-12.11.1d$ \citep{Arn96}. The errors in the values of all the best fit parameters were estimated at 90\% confidence level. The combined spectra were well represented by a thermal disk emission component and a Comptonization component. Disk emission was modelled using multi-colour blackbody emission ($\mathtt{diskbb}$), whereas the Comptonization component was fitted with Comptonization model ($\mathtt{simpl}$) \citep{Ste09}. Interstellar absorption model: Tuebingen$-$Boulder Inter Stellar Medium absorption model ($\mathtt{tbabs}$) \citep{Wil00} was used to compensate for the photon loss due to the intervening medium between the source and observer. Model $\mathtt{relxill}$ \citep{Gar14} was included to account for the relativistic reflection component from the accretion disk in the vicinity of the black hole by freezing the disk inclination angle and black hole spin, respectively, at $77.6^{\circ}$\citep{Mai14} and $0.998$. We discuss the variations of these parameters in the next section. Iron abundance (A$_{\rm{Fe}}=1.0$) , log ionisation (logX$_i~=1.0$) and high energy cut-off (E$_{\rm{cut}}=100.0$~keV) parameters were fixed during the fitting. For {$\mathtt{relxill}$}, the inner disk radius was taken to be the last stable orbit while the outer one was kept at $400$~R$_{\rm g}$, where R$_{\rm g}$ = $GM_{BH}/c^{2}$ is the gravitational radius, G is the universal gravitational constant (cm$^{3}$ g$^{-1}$ s$^{-2}$) and c is the speed of light (cm s$^{-1}$). The power law index parameters for $\mathtt{simpl}$ and $\mathtt{relxill}$ were together treated as a single variable. We introduced a constant factor for all spectral fits to account for any overall calibration uncertainty between different instruments. While the constant was fixed to unity for LAXPC, it was allowed to vary for SXT in the \textit{AstroSat} data analysis. The SXT constant was allowed to vary while those of LAXPC and XRT were fixed to unity in \textit{AstroSat/Swift} analysis. For \textit{Nustar/Swift} analysis, the variable constant was applied to XRT spectra. The best fit representative spectra and the residuals are shown in the top left (Epoch 7) and bottom left (Epoch 8) of Figure \ref{combspec} and the best fit parameters are presented in Table \ref{disk}.
\begin{table*}
\centering
   \caption{Best fit spectral parameters for the model combination $\mathtt{constant \times tbabs \times (relxill + simpl \otimes~diskbb)}$ for (i) \textit{AstroSat} data ($0.7-20.0$~keV), (ii) \textit{AstroSat and Swift} data ($0.3-20.0$~keV) and (iii) \textit{Swift and NuSTAR} data ($0.3-50.0$~keV). The best fit constant factor for \textit{ AstroSat} data and \textit{AsroSat/Swift} data was applied to the SXT spectra, while for \textit{NuSTAR/Swift}, it was applied to the XRT data.}
    \begin{tabular}{clllllllllll}
    \toprule
    \large{(i) \textit{AstroSat}} \\
    \midrule
         Epoch & constant & Hydro- & relxill & Asympto & Scattered & Tempera- & Disk & Total & Disk & Flux & Reduc-\\
         & relative & gen & norma- & tic power- & fraction of & ture at & norma-& unabs- & flux & ratio &ed chi\\
         & norma- & column & lization & law & the seed & inner disk  & lization & orbed & & & square\\
        & lization & density &  & index & photons & radius & & flux & & & \\
         &  & N$_{\rm{H}}$ & N$_{\rm{relxill}}$ & $\Gamma$ & FractSctr & kT$_{\rm{in}}$ & N$_{\rm{disk}}$ & F$_{\rm{total}}$ & F$_{\rm{disk}}$& & $\chi{^2}/$dof \\
        &  & $\times10^{22}$ & $\times10^{-4}$ & & $\times10^{-2}$& & & $\times10^{-9}$ & $\times10^{-9}$ & & \\ 
        &  & (\pcm) & & & & (keV)& & (erg\pcms) & (~erg\pcms) &(\%) & \\ 
         \midrule
         1 & 1.23$^{+0.04}_{-0.04}$ & 0.11$^{+0.02}_{-0.02}$ & 2.08$^{+1.98}_{-1.45}$ & 2.75$^{+0.36}_{-0.25}$ & $<1.41$ & 1.58$^{+0.02}_{-0.02}$ & 9.03$^{+0.77}_{-0.71}$ & 1.111$^{+0.035}_{-0.033}$ & 1.084$^{+0.029}_{-0.027}$ & 97.57 & 62$/$99 \\
         \\
         2 & 1.18$^{+0.05}_{-0.05}$ & 0.12$^{+0.01}_{-0.01}$ & $<1.06$ & 2.36$^{+0.24}_{-0.22}$ & 10.93$^{+3.94}_{-2.90}$ & 1.42$^{+0.04}_{-0.04}$ & 11.45$^{+1.58}_{-1.34}$ & 1.017$^{+0.033}_{-0.032}$ & 0.894$^{+0.029}_{-0.030}$ & 87.91 & 71$/$96 \\
         \\
        3 & 1.13$^{+0.04}_{-0.04}$ & 0.13$^{+0.03}_{-0.03}$ & 2.28$^{+2.28}_{-1.55}$ & 2.29$^{+0.32}_{-0.42}$ & 3.49$^{+2.64}_{-1.90}$ & 1.54$^{+0.04}_{-0.04}$ & 10.55$^{+1.50}_{-1.28}$ & 1.235$^{+0.049}_{-0.046}$ & 1.152$^{+0.034}_{-0.032}$ & 93.28 & 63$/$96 \\
         \\
         4 & 1.31$^{+0.04}_{-0.04}$ & 0.16$^{+0.02}_{-0.02}$ & 2.57$^{+1.75}_{-1.19}$ & 2.39$^{+0.16}_{-0.20}$ & $<0.38$ & 1.59$^{+0.02}_{-0.02}$ & 8.42$^{+0.65}_{-0.59}$ & 1.101$^{+0.035}_{-0.031}$ & 1.057$^{+0.027}_{-0.025}$ & 96.00 & 101$/$100 \\
         \\
          5 & 1.21$^{+0.04}_{-0.04}$ & 0.17$^{+0.04}_{-0.04}$ & 4.79$^{+3.36}_{-2.41}$ & 2.52$^{+0.13}_{-0.20}$ & $<0.86$ & 1.53$^{+0.02}_{-0.02}$ & 10.50$^{+1.04}_{-0.92}$ & 1.182$^{+0.047}_{-0.043}$ & 1.113$^{+0.035}_{-0.032}$ & 94.16 & 83$/$94 \\
          \\
          6 & 1.12$^{+0.05}_{-0.05}$ & 0.16$^{+0.05}_{-0.05}$ & 4.22$^{+6.52}_{-3.60}$ & 2.83$^{+0.22}_{-0.38}$ & $<1.08$ & 1.34$^{+0.03}_{-0.03}$ & 13.58$^{+1.74}_{-1.52}$ & 0.881$^{+0.048}_{-0.044}$ & 0.842$^{+0.036}_{-0.032}$ & 95.57 & 77$/$83 \\
         \bottomrule\\
        \large{(ii) \textit{Swift~$+$~}}\\\large{\textit{AstroSat}}\\
         \midrule \\
         7 & 0.81$^{+0.03}_{-0.03}$ & 0.15$^{+0.02}_{-0.02}$ & 2.79$^{+1.92}_{-1.40}$ & 2.51 $^{+0.26}_{-0.25}$ & 7.51$^{+3.07}_{-2.37}$ & 1.56$^{+0.04}_{-0.04}$ & 11.12 $^{+1.28}_{-1.09}$ & 1.531$^{+0.060}_{-0.053}$ & 1.378$^{+0.038}_{-0.036}$ & 90.01 & 106$/$148 \\
         \bottomrule\\
         \large{(iii) \textit{Swift+}}\\ \large{\textit{NuSTAR}} \\ 
         \midrule \\
         8 & 0.95$^{+0.02}_{-0.02}$ & 0.23$^{+0.02}_{-0.02}$ & 13.03$^{+3.39}_{-3.89}$ & 3.01 $^{+0.11}_{-0.09}$ & 1.52$^{+0.59}_{-0.63}$ & 1.61$^{+0.01}_{-0.01}$ & 11.72 $^{+0.31}_{-0.29}$ & 1.307$^{+0.007}_{-0.007}$ & 1.268$^{+0.007}_{-0.006}$ & 97.02 & 572$/$567 \\
         \bottomrule
    \end{tabular}
    \label{disk}
\end{table*}
\\[6pt]
The disk normalization, N$_{\rm{disk}}$ varies from $8.42^{+0.65}_{-0.59}$ for Epoch 4 to $13.58^{+1.74}_{-1.52}$ for Epoch 6 (Table \ref{disk}). The variation of N$_{\rm{disk}}$ may be due to the change in the apparent inner disk radius, r$_{\rm{in}}$ (km), 
\begin{equation}
\begin{aligned}
N_{disk} = \biggl(\frac{r_{in}}{D_{10}}\biggl)^2 cos(i)
\end{aligned}
\end{equation}
where, $D_{10}$ is the distance to the source (in units of 10 kpc) and $i$ is the disk inclination angle ($^\circ$) of the source. The absolute or true or colour corrected inner disk radius ($R_{\rm in}$) is related to r$_{\rm{in}}$ as $R_{\rm in} \simeq f_c^2\times r_{in}$, where $f_c$ is the colour hardening factor. Thus the variation of N$_{\rm{disk}}$ implies that either the inner disk radius has changed by $\sim 25$\% or the colour hardening factor is not constant and has a variation of $\sim 12$\%. Figure \ref{rin} shows the absolute inner disk radii inferred by considering the standard colour hardening factor value $f_c = 1.7$ \citep{Shi95} and inclination angle of $\sim 78^{\circ}$ \citep{Mai14} at four different distances $D_{10}$ = 5~kpc, 10~kpc, 20~kpc and 30~kpc for various epochs. The inner most stable circular orbit radii ($R_{\rm{ISCO}}$) for a non-spinning black hole of mass 5~M$_\odot$ and $10$~M$_\odot$ are  marked with the solid lines in Figure \ref{rin}. Since the inner disk radii should be larger than the $R_{\rm{ISCO}}$ values for a non-spinning black hole and for a colour factor close to the standard value, the distance must be larger than 30 kpc for a black hole of mass $>5$~M$_\odot$. The values of inner most stable circular orbit radii ($R_{\rm{ISCO}}$) for rapidly spinning black hole of mass 5~M$_\odot$, 10~M$_\odot$ and 15~M$_\odot$ are also plotted in the Figure \ref{rin}. Thus for a rapidly spinning black hole, more plausible values such as distance of $\sim 10$~kpc and black hole mass of $< 10$~M$_\odot$ are allowed. We note that the inferred luminosity of the source is around $\sim 5 \times 10^{36}$~erg\pcms and $\sim 5 \times 10^{38}$~erg\pcms for an assumed distance of 5 and 30~kpc, and hence the source is sub-Eddington if the black hole mass is $> 5$ ~M$_\odot$.
\\[6pt]
The above inferences are based on the model $\mathtt{diskbb}$, which is an approximation to the spectra arising from a relativistic disk, extending nearly to the $R_{\rm{ISCO}}$ value. In order to study the more detailed variation of physical parameters such as the accretion rate, we fit the spectra by replacing $\mathtt{diskbb}$ with the relativistic disk model $\mathtt{kerrd}$ \citep{Lao91,Ebi03}. The XSPEC model $\mathtt{kerrd}$ has been chosen instead of $\mathtt{kerrbb}$, since the latter allows for variation of the inner disk radius as a parameter and the analysis using $\mathtt{diskbb}$ suggests that the inner disk radius varies for the different Epochs (Figure \ref{rin}). Based on the analysis mentioned above we fix the spin of the black hole to $a = 0.998$ and the source distance $10$~kpc. The inner disk radii in models $\mathtt{relxill}$ and $\mathtt{kerrd}$ were tied together and treated as one variable. Outer disk radii in the models $\mathtt{relxill}$ and $\mathtt{kerrd}$ were assigned to 400~R$_{\rm g}$ and $1\times10^5$~R$_{\rm g}$, respectively. Mass accretion rate and photon index parameters were allowed to vary during the fitting. Remaining parameters were treated in the same way as described in \S~\ref{spectra}. 
\\[6pt]
We first analysed the Epoch~4 data, which has the smallest inferred inner disk radius from the $\mathtt{diskbb}$ model and assume that for this observation the inner disk radius is equal to the $R_{\rm{ISCO}}$ value of $R_{\rm in} =$ 1.235~$R_{\rm g}$. For this spectral fitting, owing to the complexity of the model and the low value of scattering fraction, the photon index ($\Gamma$) was not constrained and hence was fixed to $\sim2.4$ (i.e. the value obtained before (Table \ref{disk})). For this combination the black hole mass was constrained to be $6.10^{+0.82}_{-0.70}$~M$_{\odot}$ and the inclination angle of $i = 72.5^{+2.10}_{-2.04}$ degree. We then fitted the other observations with the mass fixed at $6.10$~M$_{\odot}$ and inclination angle of $i = 72.5^{\circ}$, but now allowing for the inner disk radius to vary. The results of the spectral fits are tabulated in Table \ref{kerrdtab}. Alternatively, we can also fix the inner disk radii for all the other observations at $R_{\rm ISCO}$ and leave the colour hardening factor to vary. This led to the variation of the colour hardening factor from $1.7$ (fixed for Epoch 4) to $1.46$.
\begin{table*}
\centering
   \caption{Best fit spectral parameters for the model combination $\mathtt{constant \times tbabs \times (relxill + simpl \otimes~kerrd)}$ for (i) \textit{AstroSat} data ($0.7-20.0$~keV), (ii) \textit{AstroSat and Swift} data ($0.3-20.0$~keV) and (iii) \textit{Swift and NuSTAR} data ($0.3-50.0$~keV), with fixed disk inclination ($72.5^{\circ}$), black hole mass ($6.10$~M$_{\odot}$) , black hole spin ($0.998$) and source distance ($10$~kpc). The best fit constant factor for \textit{ AstroSat} data and \textit{AsroSat/Swift} data was applied to the SXT spectra, while for \textit{NuSTAR/Swift} it was applied to the XRT data.}
    \begin{tabular}{clccccccccc}
    \toprule
    \toprule
    \large{(i) \textit{AstroSat}} \\
    \midrule
    Epoch & constant & Hydrogen & relxill & Asymptotic & Scattered & Accretion & Inner disk & Eddington & Reduced\\
          & relative & column & norma-& power-law & fraction of &rate & radius &  ratio & chi-square\\
         & norma-& density & lization & index &the seed & &  \\
        & lization & &  & & photons & & \\
    & & N$_{H}$ & $N_{relxill}$& $\Gamma$ &FractSctr & $\dot{M}$ & $R_{\rm in}$ & L$^{(a)}$ & $\chi{^2}/$dof  \\
    &  & $\times10^{22}$ &  $\times10^{-4}$ & &  $\times10^{-2}$ &  $\times10^{18}$ &  &\\ 
    &  & (\pcm) &  &  & &  (g\ps) & $(R_{\rm g})$ & &\\
         \midrule
         1 & 1.21$^{+0.02}_{-0.03}$ & 0.08$^{+0.03}_{-0.03}$ & 1.74$^{+1.83}_{-1.36}$ & 2.75$^{(f)}$ & $<0.38$ & 0.084 $^{+0.009}_{-0.006}$ & 1.869$^{+0.231}_{-0.288}$ & 0.054$^{+0.006}_{-0.004}$ & 79$/$100 \\
         \\
         2 & 1.16$^{+0.04}_{-0.04}$  & 0.12$^{+0.01}_{-0.01}$ & $<1.65$ & 2.32$^{+0.20}_{-0.17}$ & 9.50 $^{+3.24}_{-2.32}$ & 0.091 $^{+0.009}_{-0.008}$ & 2.511$^{+0.176}_{-0.172}$ & 0.058$^{+0.006}_{-0.005}$ & 77$/$96 \\
         \\
         3 & 1.12$^{+0.04}_{-0.03}$ & 0.13$^{+0.02}_{-0.03}$ & 2.89$^{+1.49}_{-1.47}$ & 2.29$^{(f)}$ & 2.78 $^{+1.30}_{-1.25}$ & 0.106 $^{+0.011}_{-0.009}$ & 2.320$^{+0.160}_{-0.168}$ & 0.068$^{+0.006}_{-0.007}$ & 77$/$97 \\
         \\
         4 & 1.31$^{+0.04}_{-0.04}$ & 0.15$^{+0.01}_{-0.01}$ & 3.05$^{+0.42}_{-0.42}$ & 2.39$^{(f)}$ & $<0.38$ & 0.076 $^{+0.004}_{-0.003}$ & 1.235$^{(f)}$ & 0.049$^{+0.002}_{-0.003}$ & 52$/$82 \\
         \\
         5 & 1.21$^{+0.04}_{-0.04}$ & 0.15$^{+0.01}_{-0.01}$ & $<4.90$ & 2.57$^{+0.03}_{-0.03}$ & $<0.86$ & 0.096 $^{+0.009}_{-0.009}$ & 2.208$^{+0.158}_{-0.167}$ & 0.062$^{+0.006}_{-0.006}$ & 101$/$95 \\
         \\
         6 & 1.21$^{+0.05}_{-0.05}$ & 0.19$^{+0.06}_{-0.06}$ & 9.20$^{+12.60}_{-6.28}$  & 3.05$^{+0.19}_{-0.27}$ &  $<1.08$ & 0.109 $^{+0.019}_{-0.016}$ & 3.029$^{+0.297}_{-0.272}$ & 0.070$^{+0.010}_{-0.012}$ & 87$/$84 \\
         \bottomrule\\
          \large{(ii) \textit{AstroSat~$+$~}}\\\large{\textit{Swift}}\\
         \midrule \\
         7 & 0.80$^{+0.02}_{-0.02}$ & 0.19$^{+0.02}_{-0.02}$ & 5.09$^{+2.09}_{-2.70}$ & 2.54$^{+0.19}_{-0.20}$ & 6.62 $^{+2.45}_{-2.15}$ & 0.133 $^{+0.011}_{-0.010}$ & 2.553$^{+0.155}_{-0.157}$ & 0.085$^{+0.006}_{-0.007}$ & 117$/$148 \\
         \bottomrule\\
         \large{(iii) \textit{Swift~$+$~}}\\\large{\textit{NuSTAR}}\\
         \midrule \\
         8 & 0.94$^{+0.01}_{-0.01}$ & 0.28$^{+0.02}_{-0.02}$ & 25.65$^{+8.89}_{-7.13}$ & 3.09$^{+0.10}_{-0.11}$ & 0.66$^{+0.74}_{-0.64}$ & 0.174 $^{+0.005}_{-0.005}$ & 2.729$^{+0.051}_{-0.059}$ & 0.112$^{+0.003}_{-0.003}$ & 565$/$567\\
         \bottomrule
    \end{tabular}
    \begin{flushleft}
    $^{(a)} L~=~L_{\rm acc}/L_{\rm Edd}$, where $L_{\rm acc} = \eta\dot Mc^{2}$ with $\eta\sim0.3$ for $a = 0.998$ and L$_{\rm Edd}$~=~$1.3\times10^{38}\times(M_{\rm{BH}}/M_{\rm{\odot}})$~erg\ps, $M_{\rm{BH}}~=~6.10~M_{\rm{\odot}}$ \\
        \end{flushleft}
    \label{kerrdtab}
    \end{table*}
\\[6pt]
The spectra and the residuals for Epoch 7 (which includes \textit{AstroSat} and \textit{Swift} data) and for Epoch 8 (\textit{Swift} and \textit{NuSTAR} data) are shown in the bottom left and right panel of the Figure \ref{combspec}. The primary component is the disk emission which dominates below 20.0~keV. However, note that the blurred reflection has a significant contribution below 1.0 ~keV. There is also a  correlation  between the inner disk radius and the mass accretion rate as shown in Figure \ref{parvar}. The Spearman Rank correlation tests confirmed the correlation with a coefficient $r=0.929$ and a null hypothesis probability of $p = 0.001$. 
\section{Discussion and Summary}
Spectral analysis of seven \textit{AstroSat} observations of 4U~1957+115 (one of them having simultaneous \textit{Swift} Observation) along with a joint \textit{Swift/NuSTAR} observation in the energy range $0.5-50.0$~keV was conducted. The spectra are well described by a disk emission ($\mathtt{diskbb}$), thermal Comptonization ($\mathtt{simpl}$) and a blurred reflection component ($\mathtt{relxill}$). 
The normalization of the disk emission was found to vary for the different observations, implying that the inner disk radii may have changed. Thus, a model where the disk extends to the last innermost stable circular orbit (ISCO) (i.e. $\mathtt{kerrbb}$) cannot be applied to all the observations. On the other hand, it is also possible that the variation in the disk emission normalisation is caused by changes in the colour factor. From the inferred values of the inner disk radii, a non-spinning black hole should have mass $>5$~M$_{\rm\odot} $ and at distance larger than $30$~kpc, while for a rapidly spinning black hole, these estimates turn out be more reasonable values with a black hole mass of $< 10$~M$_{\rm\odot}$ and distance $\sim10$~kpc.
\\[6pt]
The spectral model used in the initial analysis was an approximate one, since the relativistic nature of the disk emission was not taken into account. Epoch~4 had the least estimate of inner disk radius and hence we identified that data as being represented by a disk extending to the last stable orbit. Assuming a distance of 10~kpc and  a black hole spin of 0.998, we fitted Epoch~4 with a relativistic disk model and were able to constrain the disk inclination angle to $\sim72^\circ$ and a black hole mass of $\sim 6.1$~M$_{\rm\odot}$. If instead we assume some other data set such as Epoch 8 (\textit{Swift/NuSTAR} data) to be extending to the last stable orbit we get estimates of the black hole mass and inclination angle to be $\sim 4.2$~M$_{\rm\odot}$ and $\sim61^\circ$. However, we emphasise that these estimates are based on the assumption that the distance to the source is $10$~kpc and the black hole spin is $0.998$. After fitting Epoch 4, with a disk extending to the last stable orbit, we fitted the other epochs allowing for the radii to vary, but fixing the inclination and black hole mass. Significant variation of inner disk radius was found with the largest value to be $\sim$3~R$_{\rm g}$, which is nearly twice the last stable orbit (1.235~R$_{\rm g}$) assumed for Epoch 4. The relativistic disk component fitting allows to constrain the accretion rate for the different observations and we find a positive correlation between the accretion rate and inner disk radii. This is in contrast to the behaviour of black hole X-ray transients where during the beginning of the transition, the system is in the low hard state with a low accretion rate and large inner disk radius. The system then evolves over time-scales of days, with increasing accretion rate to a soft state where the inner disk radius is small. Hence for such systems, the inner disk radius is inversely correlated with the accretion rate. On the other hand, time resolved spectral analysis of rapid ($\sim$minutes) variability of black hole systems such as GRS~1915+105 \citep{Raw22}, reveal that for such systems and time-scales, the inner disk radius correlates with the accretion rate. Our results suggest that this correlation maybe extended to longer time variations of persistent black hole systems in the disk dominated (i.e. soft ) state. Long and short term monitoring observations of such persistent sources using sensitive detectors would provide more clues regarding the drivers of the spectral variability of black hole systems.
\\[6pt]
The results presented here are from several observations which span three years, each covering a wide energy range. One of the principle results of this analysis is that the inner disk radius varies in time and seems to be correlated to the accretion rate. Hence for a given observation it may not be identified with the last stable orbit, as has been done in previous studies. However, identifying the observation having the smallest inner disk radius with the last stable one, we find qualitatively similar results to what has been reported earlier which is that the source has an highly spinning black hole of mass $\sim$~6~M$_{\odot}$ and inclination angle of $\sim$~65$^{\circ}$. The uncertainty in the distance to the source does not allow for more stringent constraints.
\section*{Acknowledgements}
The authors thank the SXT and LAXPC POC teams at Tata Institute of Fundamental Research (TIFR), Mumbai, India for verifying and releasing the data via the Indian Space Science Data Centre (ISSDC) and providing the necessary software tools. This publication uses data from the AstroSat mission of the Indian Space Research Organisation (ISRO). This work has made use of data and/or software provided by National Aeronautics and Space Administration (NASA)'s  HEASARC. One of the authors (SBG) thanks the Inter-University Centre for Astronomy and Astrophysics (IUCAA), Pune, India for the Visiting Associateship. MSP thanks IUCAA, Pune for providing facilities to complete part of this work. We thank the anonymous referee for the valuable suggestions and comments, which improved the content of the manuscript further. 
\section*{Data availability}
The data utilised in this article are available at \textit{AstroSat}$-$ISSDC website (\url{http://astrobrowse.issdc.gov.in/astro_archive/archive/Home.jsp)}, \textit{NuSTAR} Archive (\url{https://heasarc.gsfc.nasa.gov/docs/nustar/nustar_archive.html}) and \textit{Swift} Archive Download Portal (\url{https://www.swift.ac.uk/swift_portal/}). The software used for data analysis is available at HEASARC website (\url{https://heasarc.gsfc.nasa.gov/lheasoft/download.html)}.
\bibliography{bibliography}{}
\bibliographystyle{\mnras}
\bsp	
\label{lastpage}
\end{document}